\documentclass{jps-cp}
\usepackage{txfonts} %Please comment out this line unless the txfonts package is availabe in your LaTeX system.
\title{Multi-messenger Astrophysics with the Pierre Auger Observatory}

\author{Michael Schimp$^{1}$ for the Pierre Auger Collaboration$^{2}$}

\inst{$^{1}$Bergische Universit\"at Wuppertal, Gau\ss{}str. 20, 42119 Wuppertal, Germany \\
$^{2}$Observatorio Pierre Auger, Av.\ San Mart\'in Norte 304, 5613 Malarg\"ue, Argentina\\ (Full author list: http://www.auger.org/archive/authors\_2019\_02.html)}

\email{\rm auger{\_}spokespersons@fnal.gov}

\recdate{\today}

\abst{While the Pierre Auger Observatory is a very successful instrument for ultra-high energy cosmic ray (UHECR) detection, it is increasingly used as part of various types of multi-messenger searches, in which it contributes with searches for air showers induced by atomic nuclei, neutrons, photons, and neutrinos.
We present an overview of the multi-messenger activities of the Pierre Auger Observatory.
The overview includes:
searches for ultra-high energy photons and neutrinos detected by the Pierre Auger Observatory in coincidence with gravitational wave events detected by LIGO and Virgo;
searches for correlations of the arrival directions of UHECRs detected by the Pierre Auger Observatory and high-energy neutrinos detected by IceCube and ANTARES;
searches for Galactic neutrons;
the multi-messenger campaign ``Deeper, Wider, Faster'', aiming for common observations of a variety of complementary instruments.
We discuss the motivations, methods and results of these searches.}

\kword{multi-messenger astrophysics, Pierre Auger Observatory, neutrinos, photons, ultra-high energy cosmic rays, neutrons  }

\begin{document}
\maketitle

\section{Introduction}
Similar to multi-wavelength observations in photon-based astronomy that have substantially broadened the general astrophysical understanding since the 1950s, multi-messenger astrophysics has already provided unique astrophysical insights that otherwise would not have been possible.
An early example is the solar storm of 1859, also known as the Carrington Event, establishing the existence of solar flares.
It was visible as very bright white light (messenger: photons) close to a set of sunspots~\cite{carrington1859description,hodgson1859curious} and additionally in the form of unusually strong auroras (messenger: cosmic rays, inducing the auroras) that have been observed even at low latitudes, for example below 9$^\circ$N in Colombia~\cite{Colombia}.

The most remarkable recent discoveries in multi-messenger astrophysics were made in the context of the detection of gravitational waves (GWs) and gamma-rays from a binary neutron star (BNS) merger~\cite{GW170817GWGR}, triggering a large search campaign for photons across a very wide range in the electromagnetic (EM) spectrum as well as searches for other messengers~\cite{GW170817MM}, consequently leading to the observation of a kilonova as a counterpart of the BNS merger~\cite{GW170817KN}.
Another so-far unique astrophysical multi-messenger observation is the detection of a high energy neutrino from the distant blazar TXS-0506+056 in coincidence with a flare of high energy photons in 2017~\cite{icecube2018highE,FermiATelFlare}.
Investigations additionally revealed a period of significantly enhanced neutrino emission from this source in 2014 and 2015~\cite{icecube2018prior}, corroborating the assumption of TXS-0506+056 being a high-energy neutrino source.

The Pierre Auger Observatory is the largest cosmic ray detector in the world, regularly used for detections of extensive air showers (EASs) induced by ultra-high-energy cosmic-rays (UHECRs), atomic nuclei roughly on an EeV energy scale.
As these are of extragalactic origin~\cite{AnisotropyScience}, they travel long distances through magnetic fields from their sources to the Earth, changing their directions, and therefore can not in general be used as precise pointers indicating their origin.

Assuming that other messengers are created at the acceleration sites of UHECRs, or during their propagation, multi-messenger astrophysical observations are a promising approach to answer the long-standing questions regarding the sites and mechanisms of ultra-high-energy particle acceleration in the universe.
In the following sections, we review various contributions of the Pierre Auger Observatory to multi-messenger astrophysical observations.
%The multi-messenger astrophysical observations described in this work are the following:
%Section~\ref{sec:GW}: Searches for ultra-high energy neutrinos with the Pierre Auger Observatory in directional and temporal coincidence with GW events provided by the LIGO/Virgo Collaboration~(LVC) are discussed.
%Section~\ref{sec:Nu} covers searches for correlations between ultra-high energy cosmic rays detected by the Pierre Auger Observatory and Telescope Array, and high-energy neutrinos detected by IceCube.

\section{Ultra-High Energy Neutrino and Photon Follow-Up Searches of LIGO/Virgo Gravitational Wave Events}~\label{sec:GW}
Searches for ultra-high energy (UHE; $>~0.1$ EeV) neutrinos and photons with the Pierre Auger Observatory have been successfully performed several times~\cite{originDiffuseNu,PSNu,photonJCAP,photonICRC2019Julian}.
The searches rely on the discrimination of EASs induced by neutrinos or photons from those induced by UHECRs, which are by far the most common EASs measured with the Pierre Auger Observatory.

The discrimination is based on the fact that these EASs develop differently in the atmosphere.
Neutrinos interact deeper in the atmosphere than the other particles, leading to more EM and hadronic particles reaching the Pierre Auger Surface Detector (SD) than in the case of UHECR-induced EASs, eventually leaving longer-lasting light signals in the photomultiplier tubes (PMTs).
This distinction is more precise and efficient for inclined showers, leading to a severely varying sensitivity across the sky~\cite{PSNu}.
Photon-induced EASs are distinguished from others based on several measures such as steeper lateral distribution functions, deeper shower maxima, and smaller footprints.
%For neutrinos, this discrimination is based on the inferred shower age, estimated from the signal detected with the photomultiplier tubes (PMTs) in the water-Cherenkov detectors of the Pierre Auger Surface Detector (SD).
%The discrimination is only efficient for high shower inclinations since highly inclined nuclei-induced EAS reach the ground with a high shower age, i.e. they have no substantial EM and hadronic component, while EAS induced by neutrinos, which interact deeper in the atmosphere, have generally a low shower age, i.e. they have a substantial EM and hadronic component even for high inclinations.
%Overall, the searches for UHE neutrinos have their combined peak sensitivity close to 1 EeV as described in detail in~\cite{originDiffuseNu} and are therefore complementary to searches by dedicated neutrino observatories like IceCube and ANTARES, which are most sensitive around the TeV range.

A multi-messenger search performed with photons and neutrinos at the Pierre Auger Observatory is the follow-up of gravitational wave (GW) events detected by the LIGO and Virgo (LV) observatories.
These were caused by the coalescence and merger of compact binaries, mostly of binary black holes (BBHs) but also of a binary neutron star (BNS).

Before VHEPA 2019, during the first two observational runs LV, called O1 and O2, GW event information was sent in form of alerts only to parties that signed a memorandum of understanding with the LV collaborations, one of which was the Pierre Auger Collaboration.
Each such alert contained the following estimated parameters if applicable:
\begin{itemize}
    \item Time of merger
    \item Masses of merged objects and remnant
    \item Distance of emitting system
    \item Sky localization probability distribution
\end{itemize}

The neutrino follow-up searches were performed by applying the default neutrino search inside the 90\% C.L. most probable localization region in the sky during a time range from 500 s before until 1 day after the merger.
No neutrino candidates have been found for any of the GW events during O1 and O2.

\begin{figure}[tbh]
\centering
\includegraphics[width=.75\textwidth]{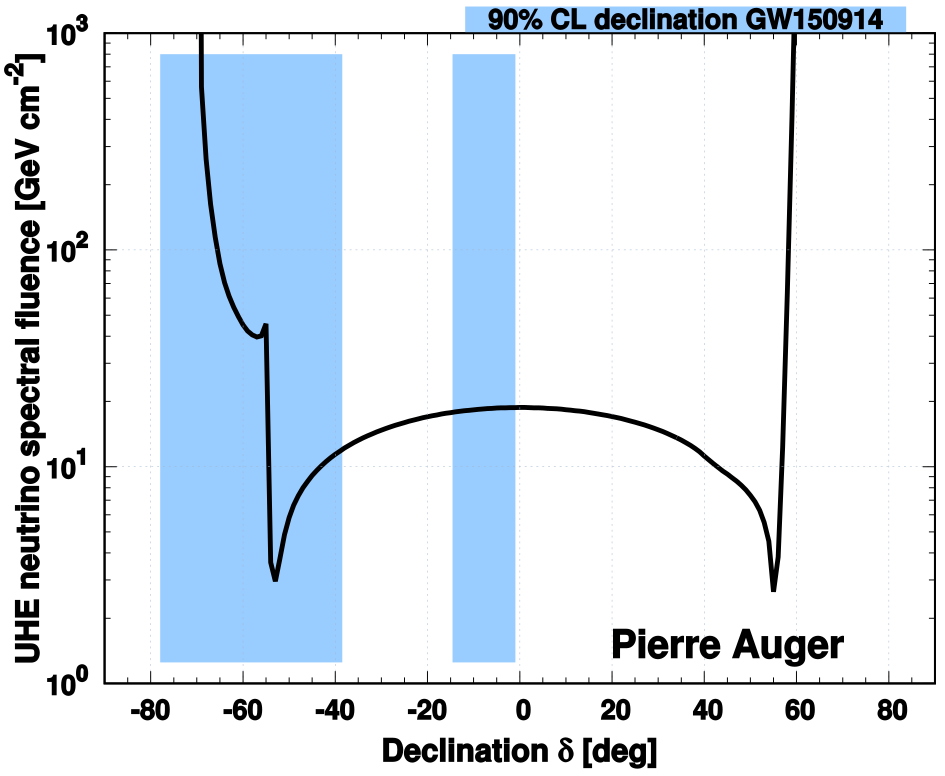}
\caption{Black lines represent the 90\% C.L. limit on the UHE neutrino fluence from GW150914 as a function of the source declination. Declinations of the 90\% C.L. sky localization of the source are highlighted in blue.~\cite{O1FU}}
\label{fig:GW150914}
\end{figure}

For each of the BBH mergers, the exposure is used to calculate limits on the UHE neutrino fluence as a function of declination of the true source, which is often known with a very limited precision (10s of degrees).
As an example, Fig.~\ref{fig:GW150914} shows the results for GW150914~\cite{O1FU}, the first GW event from a compact binary coalescence that has ever been detected~\cite{GW150914}.

\begin{figure}[tbh]
\centering
\includegraphics[width=.66\textwidth]{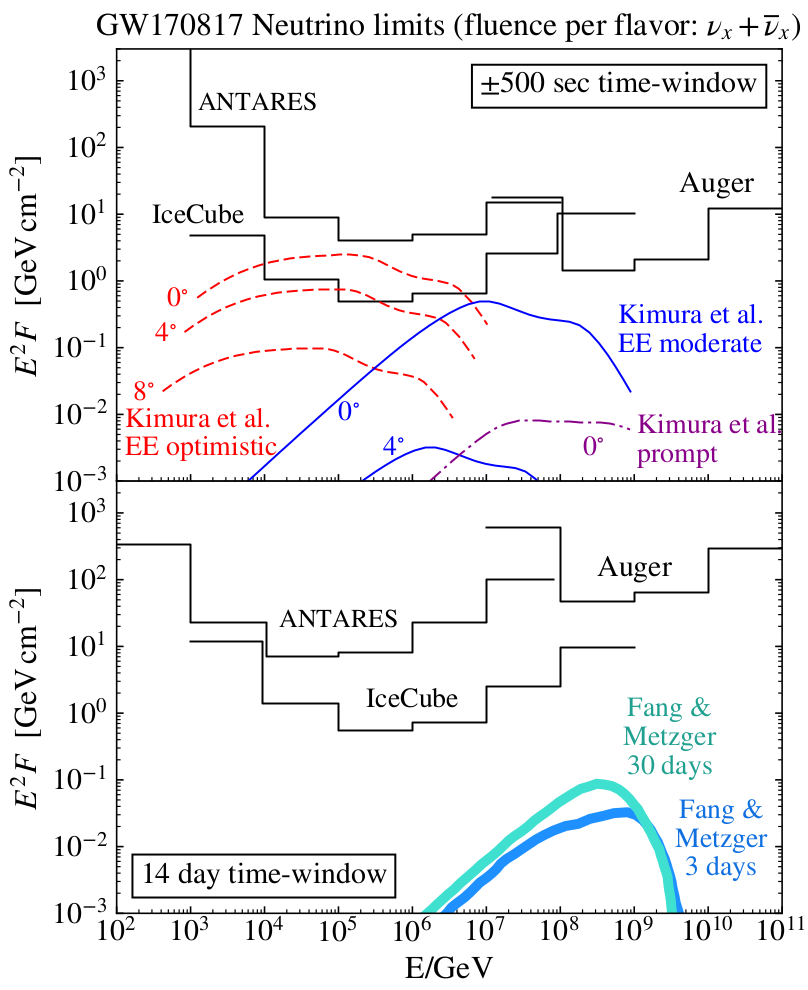}
\caption{90\% C.L. limits on the UHE neutrino spectral fluence from GW170817 as a function of energy are shown as black angular lines. Model UHE neutrino spectral fluences are represented by the colored smooth lines with the denoted off-axis angles of the merger system.~\cite{GW170817NU}}
\label{fig:GW170817}
\end{figure}

One of the GW events in O2, GW170817, has been associated with the coalescence of two objects with masses in the typical neutron star mass range ($\sim{}$1--2~$M_\odot$)~\cite{GW170817GWGR}.
Follow-up observations of this event with photons yielded signatures of a kilonova caused by a BNS merger at the inferred location of the GW event via detections in a very large wavelength range~\cite{GW170817KN, GW170817MM}:
Less than two seconds after the merger, a short GRB was observed, whereas at various lower photon energies, light curves of the source have been recorded for several weeks.
Therefore, in agreement with IceCube, the time range of the search for ultra-high energy neutrinos with the Pierre Auger Observatory from this source was extended until 14 days after the merger.
Fig.~\ref{fig:GW170817} shows the fluence limits (90\% C.L.) for the time ranges of $\pm 500$~s around the merger, and 0 to 14 days after the merger, respectively~\cite{GW170817NU}.
As the UHE neutrino sensitivity of the Pierre Auger Observatory is varying across the sky, and the BNS merger occurred at a time and location with a large sensitivity, the fluence limits for the $\pm 500$~s time range are much more competitive than for the 14 day time window, where any short-term sensitivity enhancements are averaged out due to the moving field of view of the observatory.

The photon follow-up searches are unpublished yet.
In addition to the process described for neutrinos, only GW events that are relatively close by or well localized will be taken into account in order to prevent false-positive detections.

\section{Searches for Correlations between Ultra-High Energy Cosmic Rays and High-Energy Neutrinos}~\label{sec:Nu}
UHECRs are subject to deflection in magnetic fields due to their charge, which makes finding their sources difficult.
However, as the deflection decreases with energy, correlations between particularly high-energy UHECRs  and high-energy neutrinos can be expected.
To search for these correlations, UHECRs with energies $\gtrsim{}50~\mathrm{EeV}$ detected by the Pierre Auger and the Telescope Array observatories, and high-energy neutrinos detected by IceCube, were analyzed in a joint work of the three collaborations~\cite{UHECRNUICRC2017}.

Two analyses of the correlations between these UHECRs and high-energy neutrinos have been applied: a cross-correlation analysis and a stacking likelihood analysis.
The neutrino sample contains track-like and cascade-like events which are analyzed separately.
Taking an isotropic UHECR flux as the null hypothesis, for track-like neutrino events, both analyses yielded no significant results.
However, for cascade-like neutrino events, both analyses yielded significant excesses of correlations.

\begin{figure}[tbh]
\centering
\includegraphics[width=.75\textwidth]{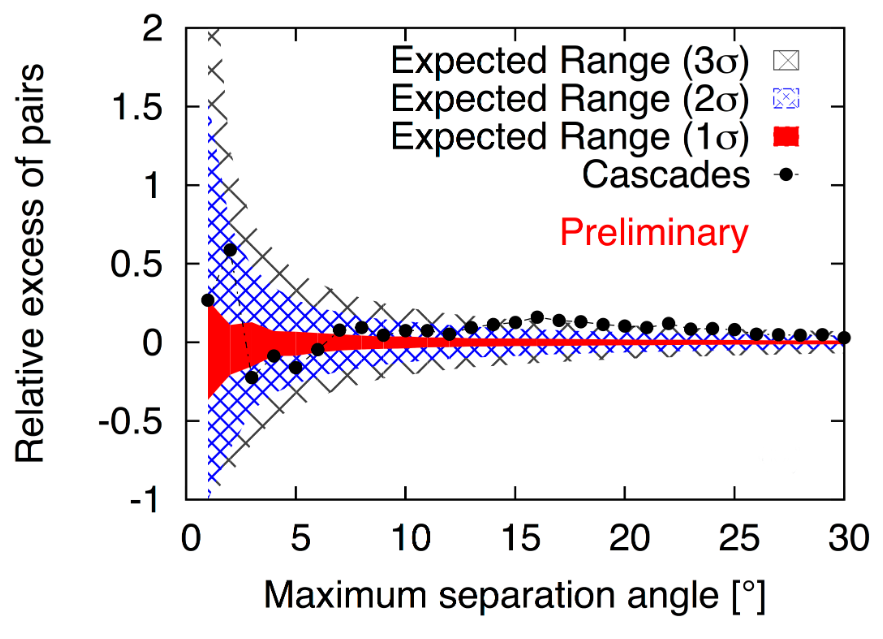}
\caption{Relative excesses of pairs as a function of the maximum angular separation are shown for the cross-correlation analysis performed with cascade-like events.
Hatched areas indicate expected fluctuations from isotropically distributed UHECRs.~\cite{UHECRNUICRC2017}}
\label{fig:UHECRNUCORR}
\end{figure}

The most significant excess in the cross-correlation analysis was found for a maximum angular separation of $22^\circ$, with a post-trial $p$-value of $5.4\cdot{}10^{-3}$.
Fig.~\ref{fig:UHECRNUCORR} shows the excess of pairs for the cross-correlation analysis as a function of the maximum angular separation~\cite{UHECRNUICRC2017}.
In the stacking likelihood analysis, the most significant excess was found for a UHECR deflection of $\frac{6^\circ}{E/100\mathrm{EeV}}$, with a post-trial $p$-value of $2.2\cdot{}10^{-2}$.
Assuming, alternatively, a null hypothesis of an isotropic high-energy neutrino flux, very similar levels of confidence for the excesses were found~\cite{UHECRNUICRC2017}.

In~\cite{UHECRNUICRC2019}, a study published after VHEPA~2019, including also the ANTARES neutrino telescope, deviations from the null hypotheses were found to be much weaker than in ~\cite{UHECRNUICRC2017}, indicating that the correlations searched for are not very strong.
This can be explained by several factors, e.g. uncertainties of the magnetic fields responsible for UHECR deflection, and the fact that UHECRs detected at Earth come from not more than a few 100 Mpc away, while possibly a large fraction of the detected neutrinos originates from much further distances.
For these neutrinos, a correlation would be unexpected, diluting the overall effect.

\section{Search for a flux of Ultra-high Energy Neutrons from the Galaxy}

Individual neutron-induced EASs are indistinguishable from those induced by protons.
However, as neutrons travel in straight lines, a flux of neutrons could be detected via an excess of the number of EASs from the directions of their sources.
As the neutron decay length is $\sim{}9.1~\mathrm{kpc}~E/\mathrm{EeV}$~\cite{PDG}, neutrons with energies of a few EeV can reach the Earth from the entire Galaxy but not from much further away.
Therefore, 11 classes of sources in the Galaxy have been used as combined target sets for neutron searches with the Pierre Auger Observatory~\cite{Neutrons}.

The chosen target sets are the Galactic center and plane as well as known photon source classes like pulsars and X-ray binaries, amounting several hundred sources in total.
The choice of photon sources as probable neutron sources is motivated by the fact that both messengers are produced in photo-hadronic interaction scenarios.

The searches in the first 9.75 years of data taken by the observatory have yielded no significant excess of the number of EASs from the target sets with respect to the neutron-free expectation.
However, from the sensitivity to the target sources, 95\% C.L. upper limits on the energy flux in neutrons of 0.1~eV~cm$^{-2}$~s$^{-1}$ (Galactic center) through 0.56~eV~cm$^{-2}$~s$^{-1}$ (Galactic plane) have been deduced for the different target sets.
In all cases, these limits already exclude energy fluxes on the level of the measured TeV photon energy flux from the target sets~\cite{Neutrons,TeVAstronomy}.

At energies of a few EeV, the UHECR composition is consistent with a large proton fraction~\cite{MassCompositionICRC2019}.
The strong limit on the neutron flux -- in particular that it is much lower than the TeV photon flux -- leads to the exclusion of an $E^{-2}$ Fermi-acceleration of protons up to energies of several EeV.
Ref.~\cite{Neutrons} also provides an interpretation in terms of an effective neutron-to-proton luminosity ratio from the Galactic plane.
For this, several theoretical assumptions regarding the proton emission and neutron production efficiency are made and related to the neutron flux limits, finally yielding a 95\% C.L. upper limit for this ratio of $0.6\%$, placing a strong constraint on ultra-high energy proton production in our Galaxy.

\section{Ultra-High Energy Cosmic Rays for the Deeper, Wider, Faster program}

The Deeper, Wider, Faster (DWF) program is a multi-instrument multi-messenger observation project.
With more than 30 associated observatories involved, the purpose of DWF is the simultaneous and common observation of sky regions with a large number of instruments at the same time, combining their complementary sensitivities.
This allows for simultaneous sensitive and wide-field measurements of multiple messengers in a variety of energy ranges~\cite{DWF}.

The observations aim for transient sources, such as fast radio bursts, which last less than a second and are therefore barely possible to follow up with instruments that are pointing in other directions and would first need to be adjusted to the region of interest.
Furthermore, the simultaneous common observation allows to observe transient sources directly before their enhanced emission, which is naturally missed by said instruments that need to adjust their pointing.
At high energies, DWF is also used to search for transients lasting seconds to hours.
Candidate identification is possible in seconds to minutes and a fast respond time of a few minutes allows for further follow-up observations with a short delay.

The Pierre Auger Observatory is contributing to DWF by sharing all detected events in the field of view of DWF.
The large field of view of the Pierre Auger Observatory lets it contribute to DWF during a large fraction of the observation time.
So far, no significant coincidences of UHECR events with other events in DWF have been found.
However, the possibility of detecting such coincidences makes DWF an interesting approach for multi-messenger observations with unprecedented combinations of messengers, including UHECRs.

%\begin{table}[tbh]
%\caption{Captions to tables and figures should be sentences.}
%\label{t1}
%\begin{tabular}{ll}
%\hline
%AAA & BBB \\
%CCC & DDD \\
%\hline
%\end{tabular}
%\end{table}

%The \verb|seceq| option resets the equation numbers at the start of each section.

%\begin{figure}[tbh]
%\includegraphics{fig01.eps}
%\caption{Captions to tables and figures should be sentences.}
%\label{f1}
%\end{figure}

%Label figures, tables, and equations appropriately using the \verb|\label| command, and use the \verb|\ref| command to cite them in the text as ``\verb|as shown in Fig. \ref{f1}|". This automatically labels the numbers in numerical order.

%The \verb|minipage| environment can be used to place figures horizontally.

%\begin{equation}
%E = mc^{2}
%\label{e1}
%\end{equation}

%\appendix
%\section{}

%Use the \verb|\appendix| command if you need an appendix(es). The \verb|\section| command should follow even though there is no title for the appendix (see above in the source of this file).

\end{document}